\documentclass[UTF8,showpacs,superscriptaddress,reprint, amsmath,amssymb, aps,twocolumn]{revtex4}
\usepackage{amsmath}
\usepackage{graphicx}
\usepackage{dcolumn}
\usepackage{bm}
\usepackage{epsfig}
\usepackage{epsf}
\usepackage{color}
\usepackage{float}
\usepackage[colorlinks=true,
urlcolor=blue,
linkcolor=blue,
anchorcolor=blue,
citecolor=blue,
]{hyperref}
\usepackage{caption}
\usepackage{subfigure}
\usepackage{multirow}
\allowdisplaybreaks[4]
\usepackage[mathlines]{lineno}

\begin{document}

\preprint{APS/123-QED}

\title{Bell nonlocality and entanglement in $\chi_{cJ}$ decays into baryon pair}

\author{Peng-Cheng Hong}
\email[]{hongpc@ihep.ac.cn}
\affiliation{College of Physics, Jilin University, Changchun 130012, People's Republic of China}
\author{Rong-Gang Ping}
\email[]{pingrg@ihep.ac.cn}
\affiliation{Institute of High Energy Physics, Beijing 100049, People's Republic of China}
\author{Wei-Min Song}
\email[]{weiminsong@jlu.edu.cn}
\affiliation{College of Physics, Jilin University, Changchun 130012, People's Republic of China}

\begin{abstract}
We present a systematic analysis of Bell nonlocality and entanglement in $\chi_{cJ}$($J=0,1,2$) decays into baryon pair($B\bar{B}$), with particular emphasis on their production via the process $e^+e^- \to \psi(2S) \to \gamma \chi_{cJ}$ at BESIII. From the baryon-antibaryon spin density matrix, we construct measurable Bell observables and concurrence, revealing a striking hierarchy of quantum correlations: $\chi_{c0}$ decays exhibit maximal violation and entanglement; $\chi_{c1}$ decays violate Bell inequalities for $\theta_1 \in (0, \pi)$ with angle-modulated strength; we find that the $B\bar{B}$ pair in $\chi_{c2}$ decays is in a separable state, and no indication of Bell inequality violation is observed. We provide complete analytical results for $J=0,1$ and quantitative, uncertainty-aware estimations for $J=2$ based on experimental inputs from BESIII. These results establish the $\chi_{cJ}$ system produced via this radiative transition as a novel and promising platform for testing quantum entanglement and Bell nonlocality in high-energy collisions.
\end{abstract}

\maketitle


\section{Introduction}\label{intro}
Quantum mechanics is fundamentally distinguished from classical theories by the existence of Bell nonlocality and quantum entanglement \cite{Horodecki:2009zz}. Tests of local realism via violations of Bell inequalities (BI) provide a definitive signature of nonclassical correlations \cite{Bell:1964kc, Clauser:1969ny}, while entanglement constitutes a key resource for quantum information \cite{Brunner:2013est}. Historically probed in low-energy systems such as photons and atoms \cite{Freedman:1972zza, Aspect:1981nv, Aspect:1982fx, Hagley:1997bob}, these quintessential quantum phenomena are now being vigorously explored in a new arena: high-energy collider experiments, where entangled particle pairs are abundantly produced. The decays of charmonium states, such as the $J/\psi(\psi(2S))$ and $\chi_{cJ}$ particles, into baryon-antibaryon pairs present a unique laboratory for such studies, combining the clarity of a quarkonium spectrum with the rich spin structure of spin-1/2 baryons.

In recent years, high-energy colliders have established themselves as a novel and powerful platform for probing quantum correlations \cite{Fabbrichesi:2021npl, Bernal:2024xhm}. The abundant production of entangled particle pairs—such as top-quark pairs at the LHC \cite{ATLAS:2023fsd} and tau-lepton pairs at Belle II \cite{Ehataht:2023zzt}—enables tests of quantum foundations at energy scales far beyond those of traditional optical and atomic experiments. As inherently relativistic systems described by the Standard Model, they offer a fundamentally new perspective on entanglement and nonlocality. Extending these studies to the hadronic sector, particularly to systems of entangled baryons, presents a unique set of opportunities and challenges.

Hyperon-antihyperon pairs produced in electron-positron annihilations constitute a particularly interesting “massive qubit” system \cite{BESIII:2018cnd, BESIII:2025vsr}. The self-analyzing nature of hyperon weak decays serves as a built-in polarimeter, enabling the full reconstruction of the spin density matrix from the angular distributions of the decay products \cite{Chung:1971ri}. This makes systems like $\Lambda\bar{\Lambda}$ and $\Xi^{+}\bar{\Xi}^{-}$ ideal candidates for quantum tomography at colliders. While most studies to date have focused on production through vector resonances such as $J/\psi$ and $\psi(3686)$ \cite{Wu:2024asu, Fabbrichesi:2024rec}, the quantum correlations in decays from scalar and tensor charmonia—the $\chi_{cJ}$ states—remain largely unexplored.

The $\chi_{cJ}$ states ($J=0,1,2$), with their distinct $J^{PC}$ quantum numbers ($0^{++}, 1^{++}, 2^{++}$), provide a rich laboratory in which the initial spin-parity structure is imprinted onto the spin correlations of the final-state baryon-antibaryon pairs. In this work, we focus on $\chi_{cJ}$ mesons produced via the radiative transition $\psi(2S) \to \gamma \chi_{cJ}$ in $e^+e^-$ annihilation, which is the primary production mechanism accessible at BESIII with high statistics. A central and open question is how these differing quantum numbers govern the violation of Bell inequalities and the degree of entanglement. This work presents a comprehensive analysis of Bell nonlocality and spin entanglement in $\chi_{cJ} \to B\bar{B}$ decays, conducted under the well-defined production condition established here.

Our analysis reveals a striking hierarchy of nonlocality across the $\chi_{cJ}$ states. We find that $\chi_{c0}$ decays maximally violate the Bell inequality. In contrast, $\chi_{c1}$ decays violate the inequality throughout the range $\theta_1 \in (0, \pi)$, with the violation vanishing only at the exact forward and backward directions ($\theta_1 = 0, \pi$). The case of $\chi_{c2}$ is the most intricate: the $B\bar{B}$ pair is found to be in a separable state, and no indication of Bell inequality violation is observed. These pronounced differences, directly rooted in the $J^{PC}$ quantum numbers of the parent $\chi_{cJ}$, yield clear and testable predictions for experiment.

The remainder of this paper is organized as follows. Section II introduces the theoretical framework for the joint spin density matrix. In Section III, we present the derived polarizations and spin correlations for the $B\bar{B}$ system in $\chi_{cJ}$ decays. Sections IV and V are devoted to the analysis of Bell nonlocality and quantum entanglement, respectively, where we provide analytical results and discuss their physical origins. Section VI presents a numerical analysis of the predicted quantum correlations using experimental inputs, followed by a detailed discussion of the experimental feasibility and future prospects for testing our predictions at BESIII and other facilities. We conclude with a summary and outlook in Section VII.

\section{Theoretical Framework}\label{TF}
\subsection{Joint spin density matrix and quantum correlations}
To analyze Bell nonlocality and entanglement in $\chi_{cJ} \to B\bar{B}$ decays, one can construct the joint spin density matrix $\rho^{B\bar{B}}$ of the baryon-antibaryon system. This matrix fully encodes all spin correlations of the $B\bar{B}$ pairs. The decay proceeds via the strong interactions, and conserves the parity. The quantum state of this composite system is completely described by its joint spin density matrix, $\rho^{B\bar{B}}$. A general and powerful parameterization in the basis of Pauli matrices is given by
\begin{align}\label{rbb}
	\begin{split}
		\rho^{B\bar{B}} &= \frac{1}{4}[I \otimes I + \sum_{i=1}^3 P_i ( \sigma_i \otimes I ) + \sum_{j=1}^3\bar{ P}_j (I \otimes \sigma_j) \\
		&\quad+ \sum_{i,j=1}^3 C_{ij} ( \sigma_i \otimes \sigma_j ) ].
	\end{split}
\end{align}
Here, $P_i$ and $\bar P_j$ are the components of the polarization vectors for $B$ and $\bar{B}$, respectively, quantifying their individual spin orientations. The real matrix $C_{i,j}$ is the spin correlation tensor, which encapsulates all information about the spin-spin correlations between the two particles. These correlations constitute the direct physical observables that give rise to both Bell nonlocality and quantum entanglement.

\subsection{Helicity formalism and the spin density matrix of $\chi_{cJ}$}
To derive the joint density matrix $\rho^{B\bar{B}}$ from the initial $\chi_{cJ}$ state, we employ the helicity formalism, which provides a natural framework for describing sequential decays and establishes a direct connection between theoretical calculations and experimental observables. A crucial simplification arises when the spin quantization axis is chosen along a particle's momentum direction; in this case, the helicity eigenstates coincide with the spin projection eigenstates.

Our derivation begins with the production of the $\psi(2S)$ vector meson in $e^+e^-$ annihilation. The spin density matrix of the $\psi(2S)$, produced via a virtual photon, takes the well-known form $\rho^{\psi(2S)} = \frac{1}{2} \mathrm{diag}[1,0,1]$. We then consider the subsequent electromagnetic radiative transition $\psi(2S) \to \gamma \chi_{cJ}$.

If the dynamics of this transition is assumed by electric dipole (E1) multipole amplitudes, which within the helicity framework one has the standard relations \cite{Karl:1975jp}, $A^1_{1,1} = A^1_{0,1},
	A^2_{2,1} = \sqrt{2} A^2_{1,1} = \sqrt{6} A^2_{0,1}$.
Here, $A^J_{\lambda_1,\lambda_2}$ denotes the helicity amplitude for the process, with $J$ the spin of the $\chi_{cJ}$, and $\lambda_1$ ($\lambda_2$) the helicity of the $\chi_{cJ}$ and photon. The first relation holds for $\psi(2S) \to \gamma \chi_{c1}$, and the second for $\psi(2S) \to \gamma \chi_{c2}$. However, the recent BESIII measurement implies that the E1 assumption is invalid for the $\chi_{c2}$ decay \cite{BESIII:2025gof}. Thus we use the measured helicity amplitude relations to construct the spin density matrix $\rho^{\chi_{cJ}}$ of the $\chi_{cJ}$ in the $\psi(2S)$ rest frame. This matrix explicitly depends on the $\chi_{cJ}$ emission angle $\theta_0$.

\begin{figure}[htbp]  
    \centering  
    \includegraphics[width=0.4\textwidth]{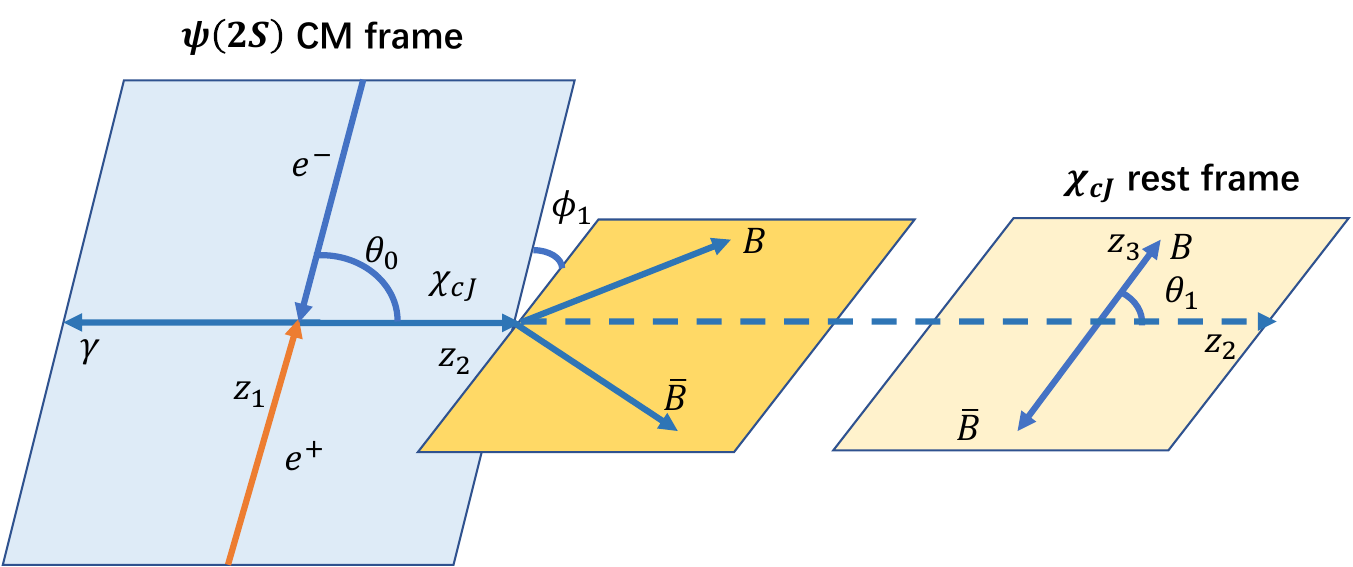}      
    \caption{The decays of $J/\psi \to \gamma \chi_{cJ}$ and $\chi_{cJ} \to B\bar{B}$}  
    \label{fig:decays} 
\end{figure}

The complete decay chain and the definitions of the relevant helicity amplitudes and angles are illustrated in Fig.~\ref{fig:decays} and summarized in Table~\ref{AmpAng}.

\begin{table}[htbp]
	\renewcommand\arraystretch{1.5}
	\caption{Helicity amplitudes and angles in the decays.}
	\label{AmpAng}
	\begin{tabular}{lll}
		\hline\hline
		decays & helicity amplitudes & helicity angles \\
		\hline
		$\psi(2S)(\lambda) \to  \chi_{cJ}(\lambda_1) \gamma(\lambda_2)$ & $A^J_{\lambda_1,\lambda_2}$ & $\Omega_1(\theta_0,\phi_0)$ \\
		$\chi_{cJ} \to B\bar{B}$ & $B^J_{\lambda_3,\lambda_4}$ & $\Omega_2(\theta_1,\phi_1)$ \\
		\hline\hline
	\end{tabular}
\end{table}

The coherences present in the initial $\chi_{cJ}$ polarization state directly get into the dynamics of the subsequent strong decay, $\chi_{cJ} \to B\bar{B}$. This process is described in the helicity formalism by amplitudes $B^{J}_{\lambda_3, \lambda_4}$, where $J$ is the spin of $\chi_{cJ}$, $\lambda_3$ and $\lambda_4$ are the helicities of the final-state baryon $B$ and antibaryon $\bar{B}$, respectively.

Conservation of $P$ and $C$ parity severely restricts the form of these amplitudes. The general constraints under P parity conservation for a decay $\chi_{cJ} \to B\bar{B}$ are given by:
\begin{align}\label{helCP}
	B^J_{\lambda_3,\lambda_4} &=(-1)^{J} B^J_{-\lambda_3,-\lambda_4},
\end{align}
where $J$ is the spins of the $\chi_{cJ}$. While the conservation of C parity requires that  \cite{Tabakin:1985yv}
\begin{align}\label{helCP}
	B^J_{\lambda_3,\lambda_4} &=(-1)^{J} B^J_{\lambda_4,\lambda_3}.
 \end{align}
This leads to the same  constraints as the parity conservation for $\chi_{c0,2}$ decays, however, it leads to the $B^1_{1/2,1/2}=B^1_{-1/2,-1/2}=0$ for $\chi_{c1}\to B\bar B$.

Applying these constraints to the specific cases of $\chi_{cJ} \to B\bar{B}$ decays yields a minimal set of independent helicity amplitudes. For $\chi_{c0}$ and $\chi_{c1}$ decays, only one independent amplitude survives: $B^0_{1/2,1/2}$ for $\chi_{c0}$ and $B^1_{1/2,-1/2}$ for $\chi_{c1}$. In contrast, the $\chi_{c2}$ decay involves two independent amplitudes. The non-vanishing amplitudes obey the following relations:
\begin{align}\label{helcj}
	B^0_{\frac{1}{2},\frac{1}{2}} &=  B^0_{-\frac{1}{2},-\frac{1}{2}},~~~
	B^{1}_{\frac{1}{2},-\frac{1}{2}} =  -B^1_{-\frac{1}{2},\frac{1}{2}},\nonumber\\
	B^{2}_{\frac{1}{2},\frac{1}{2}} &=  B^2_{-\frac{1}{2},-\frac{1}{2}},~~~
	B^{2}_{\frac{1}{2},-\frac{1}{2}} =  B^2_{-\frac{1}{2},\frac{1}{2}}.
\end{align}

For the $\chi_{c2}$ decay, the two independent amplitudes are most insightfully parameterized by their relative magnitude $x$ and phase difference $\Delta\Phi$:
\begin{align}\label{DefxPhi}
B^2_{-\frac{1}{2},\frac{1}{2}} &= x\cdot e^{i\Delta \Phi} B^2_{\frac{1}{2},\frac{1}{2}}.
\end{align}
This parameterization directly connects the decay dynamics to the observable spin correlations in the final $B\bar{B}$ system.

The elements of the spin density matrix $\rho^{\chi_{cJ}}$ are given by
\begin{align}\label{rhocj}
\rho^{\chi_{cJ}}_{\lambda_{1}, \lambda_{1}'} &\propto \sum_{\lambda, \lambda', \lambda_{1}} \rho^{\psi(2S)}_{\lambda, \lambda'} D^{1*}_{\lambda, \lambda_{1} - \lambda_{2}}(\Omega_{1}) D^{1}_{\lambda', \lambda_{1}-\lambda_{2}'} (\Omega_{1})\nonumber\\
&\times A^{1}_{\lambda_{1}, \lambda_{2}}A^{1*}_{\lambda_{1}, \lambda_{2}'}
\end{align}
where $\Omega_{1}(\theta_{0}, \phi_{0})$ is the helicity angles describing the emission angle of $\chi_{cJ}$ in $e^{+}e^{-}$ center-of-mass(CM) frame, and the Wigner $D$-function performs the necessary rotational transformation from the $e^{+}e^{-}$ rest frame (with its quantization axis defined by the $e^{+}$ momentum direction as shown in Fig.~\ref{fig:decays}) to the $\chi_{cJ}$ CM frame (where the natural quantization axis is aligned along the $\chi_{cJ}$ momentum).

The matrix $\rho^{\chi_{cJ}}$ obtained from Eq.~(\ref{rhocj}) is expressed in the $\psi(2S)$ rest frame. However, the natural quantization axis for describing the spin state of the $\chi_{cJ}$ is along its own momentum in its rest frame. To obtain the spin density matrix $\rho^{\chi_{cJ}}$ in this intrinsic frame, we must integrate over the production angle $\theta_0$ from $0$ to $\pi$. This integration removes the extrinsic kinematic dependence inherited from the $\psi(2S)$ decay and yields the intrinsically averaged polarization state of the $\chi_{cJ}$ ensemble. We note that this integrated matrix is not generally diagonal. This is especially evident for higher-spin states such as the $\chi_{c2}$ ($J=2$), where the non-diagonal elements of $\rho^{\chi_{c2}}$ directly represent coherent superpositions among different magnetic substates. 

After integrating over $\theta_0$ (and setting $\phi_0=0$ as it is always permissible), one obtains the $\rho^{\chi_{cJ}}$ in the $\chi_{cJ}$ rest frame as 
\begin{align}\label{rc0}
\rho^{\chi_{c0}} = 1,
\end{align}
\begin{align}\label{rc1}
\rho^{\chi_{c1}} ={1\over 2(1+r_1^2)} 
\begin{pmatrix}
	 r_1^2 & 0 & 0 \\
	 0 & 2 & 0 \\
	 0 & 0 & r_1^2  \\
	\end{pmatrix},
\end{align}

\begin{footnotesize}
\begin{align}\label{rc2}
\rho^{\chi_{c2}} ={1\over 4(1+r_2^2+r_3^2)}
	\begin{pmatrix}
 	2r_3^2 & 0 & r_3 e^{i\Delta \Phi_3} & 0 & 0 \\
 	0 & 2r_2^2 & 0 & 0 & 0 \\
	  r_3 e^{-i \Delta \Phi_3} & 0 & 4 & 0 & r_3 e^{-i \Delta \Phi_3} \\
 	0 & 0 & 0 & 2r_2^2 & 0 \\
 	0 & 0 & r_3 e^{i \Delta \Phi_3} & 0 & 2r_3^2 \\
	\end{pmatrix},
\end{align}
\end{footnotesize}
where $r_i$ and $\Delta\Phi_i$ are defined as the ratio of helicity amplitude,  namely, $A^1_{1,1}/A^1_{1,0}=r_1e^{i\Delta\Phi_1}$ for $\chi_{c1}$ states, while $A^2_{1,1}/A^2_{1,0}=r_2e^{i\Delta\Phi_2}$ and $A^2_{1,2}/A^2_{1,0}=r_3e^{i\Delta\Phi_3}$ for $\chi_{c2}$ state.  In the approximation of $
E1$ transition, one has $r_1=1, r_2=\sqrt 3,r_3=\sqrt 6 $  and $\Delta\Phi_i=0(i=1,2,3)$.  The recent measurement shows that the $r_2$ and $r_3$  significantly deviate from the $E1$ transition \cite{BESIII:2025gof}.

 The spin density matrix of \(\chi_{c0}\) consists of a single element and is normalized to 1, reflecting its spin-0 nature. 
The obtained spin density matrix for $\chi_{c1}$ is diagonal. This diagonal structure implies the absence of coherence among its magnetic substates ($m = \pm1, 0$) after the angular integration in the production process.
In contrast, the spin density matrix for $\chi_{c2}$ possesses non-zero off-diagonal elements (specifically between the $m = \pm2$ and $m = 0$ substates). These off-diagonal coherences are a direct consequence of the tensor polarization induced by the radiative transition from the vector $\psi(2S)$ and persist after angular integration. The presence of these coherences (e.g., $\rho^{\chi_{c2}}_{2,0} \neq 0$) is a hallmark of tensor polarization and will play a decisive role in generating the complex correlation patterns for $\chi_{c2}$ decays.

\subsection{Construction of the spin density matrix of $B\bar{B}$}
The final step in the theoretical construction is the strong decay $\chi_{cJ} \to B\bar{B}$. This decay is fully characterized by the helicity amplitudes $B^J_{\lambda_3,\lambda_4}$ and the helicity angles $\Omega_2(\theta_1,\phi_1)$.
The joint spin density matrix $\rho^{B\bar{B}}$ for the $B\bar{B}$ pair is constructed by combining the polarization information of the parent $\chi_{cJ}$, encoded in $\rho^{\chi_{cJ}}$, with the dynamical and kinematic information of the decay. The general expression reads:
\begin{align}\label{rbbele01}
	\rho^{B\bar{B}}_{\lambda_3, \lambda_4, \lambda_3', \lambda_4'} &\propto \sum_{\lambda_1,\lambda_1'} \rho^{\chi_{cJ}}_{\lambda_1,\lambda_1'}B^J_{\lambda_3,\lambda_4}B^{J*}_{\lambda_3',\lambda_4'}D^{J*}_{\lambda_1,\lambda_3-\lambda_4}(\Omega_2)\nonumber\\
	&\times D^{J}_{\lambda_1', \lambda_3^{'}-\lambda_4^{'}}(\Omega_2),
\end{align}
where $\Omega_2(\theta_1,\phi_1)$ specifies the direction of the baryon momentum in the $\chi_{cJ}$ rest frame and the angle between the $\chi_{cJ}$ production plane and its decay plane. This formalism consistently merges the decay dynamics with the relevant kinematic rotations, yielding the final, observable joint spin density matrix $\rho^{B\bar{B}}$.

\subsection{Angular distributions}\label{subsec:AngDist}
The construction of the joint spin density matrix $\rho^{B\bar{B}}$ not only provides the complete spin information for quantum correlation studies but also yields, as a direct byproduct, the differential angular distribution of the  baryon in the $\chi_{cJ}$ rest frame. 
The distribution $\mathcal{W}^J(\theta_1)$ is obtained from the trace of $\rho^{B\bar{B}}$, where $J$ is the spin of $\chi_{cJ}$.
Applying this general formula to each $\chi_{cJ}$ state, we obtain the angular distributions as follows
\begin{align}\label{eq:angdis}
\mathcal{W}^{0} &= 1, \nonumber\\
\mathcal{W}^{1} &\propto 1 + \frac{r_1^2-2}{r_1^2+2} \cos^{2} \theta_{1},\nonumber \\
\mathcal{W}^{2} &\propto 1+\frac{-6(2+r_3^2-2x^2+r_2^2(x^2-2))}{2+3r_3^2+2r_3^2 x^2 + 2 r_2^2 x^2} \cos ^2 \theta_1\nonumber\\
&+\frac{(6+r_3^2-4r_2^2)(3-2x^2)}{2+3r_3^2+2r_3^2 x^2 +2 r_2^2 x^2}\cos^4\theta_1 , 
\end{align}
for $\chi_{c0,1,2}$ states respectively.  

These analytical expressions serve as crucial benchmarks for experimental analysis. A fit to the measured angular distributions, particularly for $\chi_{c2} \to B\bar{B}$, can be used to extract or constrain the helicity amplitude parameters $x$, thereby providing independent input for the predictions of quantum correlations presented in the following sections. These distributions will be used in Sec.~\ref{NumAna} to discuss experimental constraints. We now turn to the primary focus of this work: extracting measures of quantum correlations from $\rho^{B\bar{B}}$.

\section{The Polarizations and Spin Correlations in $\chi_{cJ} \to B\bar{B}$ Decays}\label{secSpinCorr}
This section presents the core results of calculation for the joint spin density matrices for the $B\bar{B}$ system from $\chi_{cJ}$ decays. The polarization parameters $P_i^\pm$ and the spin correlation coefficients $C_{ij}$, defined in Eq.~(\ref{rbb}), are computed for each $\chi_{cJ}$ state. These results form the foundation for the analysis of Bell nonlocality and quantum entanglement in the subsequent sections.

\subsection{$\chi_{c0} \to B\bar{B}$}
The joint spin density matrix for the $B\bar{B}$ pair from the scalar $\chi_{c0}$ decay is characterized by its simplicity, reflecting the isotropic nature of the $0^{++}$ state. The $B\bar B$  is in Bell state $|\psi_+\rangle_h ={1\over\sqrt 2}(|\uparrow\uparrow\rangle + |\downarrow\downarrow\rangle )$ \footnotemark[1] .
 \footnotetext[1]{This corresponds to that in the spin basis $\protect|\psi_+\rangle_s ={1\over\sqrt 2}(|\uparrow\downarrow\rangle + |\downarrow\uparrow\rangle )$.}
Here up and down arrows denote the helicity value $\pm {1\over 2}$. Either from Eq.~(\ref{rbbele01}) or $\rho^{B\bar{B}}=|\psi_+\rangle \langle \psi_+|$ , the joint spin density matrix of $B\bar{B}$ is calculated in the helicity basis ($|\uparrow\uparrow\rangle,~|\uparrow\downarrow\rangle,~|\downarrow\uparrow\rangle~|\downarrow\downarrow\rangle,$ ) as
\begin{align}\label{rbb01}
	\rho^{B\bar{B}} =\frac{1}{2}
	\begin{pmatrix}
		1 & 0 & 0 & 1 \\
		0 & 0 & 0 & 0 \\
		0 & 0 & 0 & 0 \\
		1 & 0 & 0 & 1 
	\end{pmatrix}.
\end{align}
Comparing with the general parameterization in Eq.~(\ref{rbb}), we obtain the calculated polarization vectors and spin correlations are
\begin{align}
	P_i &= \bar P_i =  0 ~(i=x,y,z),
\end{align}
and 
\begin{align}
	C=
	\begin{pmatrix}
		1 & 0 & 0 \\
		0 & -1 & 0 \\
		0 & 0 & 1 
	\end{pmatrix}.
\end{align}

The vanishing polarizations for both $B$ and $\bar{B}$ are mandated by parity conservation in the strong decay of a $0^{++}$ state.
This pattern, $C_{xx}=-C_{yy} = C_{zz}$, is a characteristic of the spin triplet state, which  can be understood as the joint projection measurement. Alice and Bob jointly share a Bell state $|\psi_+\rangle$, and they make a correlation measurement with the operator $P^{\pm}_{{\bf n}}={1\over 2}(I\pm {\bf n}\cdot \sigma)$, and $C_{ii}=(P^+_i-P^-_i)\otimes (P^+_i-P^-_i) $ with $i=x,y$  and $z$.  For the pure states, the entanglement of $B\bar B$ can be described by the von Neumann entropy of their partial traces, i.e.,  $\rho_A=\text{Tr}_B(\rho^{B\bar{B}})=\text{diag}(1/2,1/2)$, and $H=-\sum_{i} p_i\text{log}_2(p_i)=1$. The entanglement in $\chi_{cJ}$ decays will be uniformly described using the quantum concurrence in Sections. \ref{secBell} and Sec. \ref{secCon}, respectively.  This case indicates that the $B\bar{B}$ pair from the $\chi_{c0}$ decay is produced in a pure, maximally entangled quantum state.

\subsection{$\chi_{c1} \to B\bar{B}$}
The decay of $\chi_{c1} \to B\bar{B}$ exhibits a very interesting helicity selection rule. Although the helicity states of the final-state baryons belong to a spin-1 triplet, due to the conservation requirement of $C$-parity, only the helicity state with $B^1_{1/2,-1/2} = -B^1_{-1/2,1/2}$ contributes. Therefore, only the state with total spin $0$ exists, i.e.,
\[
|\phi_{1,0}\rangle \propto B^1_{1/2,-1/2}\bigl(|\uparrow\rangle|\downarrow\rangle-|\downarrow\rangle|\uparrow\rangle\bigr).
\]
Except for a normalization factor $B_{1/2,-1/2}$, its helicity state is exactly the same as the spin singlet. This brings great convenience to our calculation.

According to Eq.~(\ref{rbbele01}), the joint spin density matrix of $B\bar{B}$ is calculated as
\begin{align}\label{rbbc1}
	\rho^{B\bar{B}} = 
	\begin{pmatrix}
	 0 & 0 & 0 & 0 \\
	 0 & \frac{1}{2} & \frac{(2-r_1^2)\sin^2 \theta_1}{2+3r_1^2+(r_1^2-2)\cos 2\theta_1} & 0 \\
	 0 & \frac{(2-r_1^2)\sin^2 \theta_1}{2+3r_1^2+(r_1^2-2)\cos 2\theta_1}  & \frac{1}{2} & 0 \\
	 0 & 0 & 0 & 0 \\
	\end{pmatrix}.
\end{align}

Here the parameter $r_1$ comes from the $\rho^{\chi_{c1}}$.  In the $E1$ transition assumption,  one has $r_1=1$.  The baryon and antibaryon polarization can be calculated with $P_i$= Tr($\rho^{B\bar B}  \sigma_i\otimes I$) and $\bar P_i$= Tr($\rho^{B\bar B}  I\otimes\sigma_i$), respectively, and  we find:
\begin{align}
	P_i &=\bar P_i=0~(i=x,y,z),
\end{align}
and the correlation parameters can be calculated with $C_{ij}= \text{Tr}(\rho^{B\bar B}\sigma_i\otimes\sigma_j)$,  then one has 
\begin{align}
	C=
	\begin{pmatrix}
		 \frac{2(2-r_1^2) \sin ^2 \theta_1}{2+3r_1^2+(r_1^2-2)\cos 2 \theta_1} & 0 & 0 \\
		0 &  \frac{2(2-r_1^2) \sin ^2 \theta_1}{2+3r_1^2+(r_1^2-2)\cos 2 \theta_1} & 0 \\
		0 & 0 & -1 
	\end{pmatrix}.
\end{align}

The structure of the spin density matrix \(\rho^{B\bar{B}}\) clearly shows how C-parity conservation restricts helicity choices in \(\chi_{c1} \to B\bar{B}\) decay. Non-zero spin density matrix elements appear only for the helicity basis states \(|\uparrow\rangle|\downarrow\rangle\) and \(|\downarrow\rangle|\uparrow\rangle\). The elements for \(|\uparrow\rangle|\uparrow\rangle\) and \(|\downarrow\rangle|\downarrow\rangle\) are all zero. The off-diagonal elements contain terms that depend on \(\theta_1\), coming from the mother particle having spin 1. This is different from the spin 0 case. For spin correlation, \(C_{zz} = -1\). This means Alice and Bob measure opposite helicities for the particles. It is a direct result of the helicity selection rule.

One key point to emphasize is this: In the decay of \(\chi_{c1}\), the helicity selection rule forces the final-state baryon spins into a Bell-state singlet form. However, compared to the singlet produced from a spin-0 mother particle, its spin density matrix has very different properties. The standard Bell-state singlet is a pure state. Its density matrix satisfies \(\text{Tr}(\rho^2)=1\). But for \(\chi_{c1}\) decay, we find \(\text{Tr}(\rho^2)<1\). This shows the \(B\bar B\) system is in a mixed state. Therefore, its quantum concurrence is less than 1. We will discuss this in detail in the next section.

\subsection{$\chi_{c2} \to B\bar{B}$}
Using this $\chi_{c2}$ density matrix in Eq. (\ref{rc2}), we derive the joint spin density matrix $\rho^{B\bar{B}}$ for the baryon-antibaryon system. 
Its structure is given by 
\begin{align}\label{rbbc2}
	\rho^{B\bar{B}} = 
	\renewcommand\arraystretch{2.0}
	\begin{pmatrix}
 	\frac{\mathcal{F}_4}{2 \mathcal{F}_1} & -\frac{\mathcal{F}_2 e^{-i \Delta \Phi }}{4 \mathcal{F}_1} & \frac{\mathcal{F}_2 e^{-i\Delta \Phi }}{4 \mathcal{F}_1} & \frac{\mathcal{F}_4}{2 \mathcal{F}_1} \\
 	-\frac{\mathcal{F}_2 e^{i \Delta \Phi }}{4 \mathcal{F}_1} & \frac{\mathcal{F}_1-\mathcal{F}_4}{2 \mathcal{F}_1} & \frac{\mathcal{F}_3}{2\mathcal{F}_1} & -\frac{\mathcal{F}_2 e^{i\Delta \Phi }}{4 \mathcal{F}_1} \\
 	\frac{\mathcal{F}_2 e^{i\Delta \Phi }}{4 \mathcal{F}_1} & \frac{\mathcal{F}_3}{2\mathcal{F}_1} & \frac{\mathcal{F}_1-\mathcal{F}_4}{2 \mathcal{F}_1} & \frac{\mathcal{F}_2 e^{i \Delta \Phi }}{4 \mathcal{F}_1} \\
 	\frac{\mathcal{F}_4}{2 \mathcal{F}_1} & -\frac{\mathcal{F}_2 e^{-i\Delta \Phi }}{4 \mathcal{F}_1} & \frac{\mathcal{F}_2 e^{-i \Delta \Phi }}{4 \mathcal{F}_1} & \frac{\mathcal{F}_4}{2 \mathcal{F}_1} \\	
	\end{pmatrix},
\end{align}
where the $\theta_1$- and $x$-dependent functions $\mathcal{F}_1$ to $\mathcal{F}_4$ are given by
\begin{align}\label{eq:F1F4}
	\mathcal{F}_1 &= (1+3\cos 2 \theta_1)^2 + 6r_3^2 \sin^4 \theta_1  \nonumber\\
 &+ 2x^2(r_3^2(3+\cos 2 \theta_1) \sin^2 \theta_1 \nonumber\\
 &+ r_2^2(2+  \cos 2\theta_1 + \cos 4\theta_1)) \nonumber\\
 &+ 6(r_2^2+x^2) \sin^2 2\theta_1, \nonumber\\
	\mathcal{F}_2 &= 2 \sqrt{6} x \sin 2 \theta_1 (r_3^2 \sin 2 \theta _1+(2r_2^2 -3) \cos 2 \theta _1-1), \nonumber\\
	\mathcal{F}_3 &= -6x^2 \sin^2 2\theta_1 + 4 r_3^2 x^2\sin^4 \theta_1 \nonumber\\
 &+ 4r_2^2x^2(1 + 2 \cos 2 \theta_1) \sin^2 \theta_1, \nonumber\\
	\mathcal{F}_4 &= (1+3\cos 2\theta_1)^2 + 6r_3^2 \sin^4 \theta_1 + 6 r_2^2 \sin^2 2\theta_1.	
\end{align}
Here  $\mathcal{F}_1$ is the normalization factor, corresponding to the unpolarized cross section.

The  polarization vectors of baryons are calculated to be
\begin{align}\label{Polc2}
	P_i &= \bar{P}_i = 0~(i=x,z), \nonumber\\
	P_y & = -\bar{P}_y =  \frac{\mathcal{F}_2 \sin \Delta \Phi }{\mathcal{F}_1}.
\end{align}

Measuring \(P_y\) depends on the interference between helicity amplitudes \(B_{1/2,1/2}\) and \(B_{-1/2,1/2}\).  
The interference strength is proportional to the sine of their phase difference.  
After integrating over \(\theta_1\) in the full phase space, we obtain the net transverse polarization \(P_y\).  
This essentially comes from the polarization transfer from \(\chi_{c2}\).  
The polarization of \(\chi_{c2}\) originates from the electromagnetic transition process of \(\psi(2S)\):  
\(\chi_{c2}\) becomes polarized by coupling with tensor-polarized photons.

The expressions for $C_{ij}$ are given as
\begin{align}
	C=\begin{pmatrix}
 	\frac{\mathcal{F}_3+ \mathcal{F}_4}{\mathcal{F}_1} & 0 & \frac{\mathcal{F}_2 \cos \Delta \Phi }{\mathcal{F}_1} \\
 	0 & \frac{\mathcal{F}_3- \mathcal{F}_4}{\mathcal{F}_1} & 0 \\
 	-\frac{\mathcal{F}_2 \cos \Delta \Phi }{\mathcal{F}_1} & 0 & \frac{2 \mathcal{F}_4- \mathcal{F}_1}{\mathcal{F}_1}
\end{pmatrix}.
\end{align}
Unlike $\chi_{c0,1}$,  the $\chi_{c2}$ particle has more spin states.  
This makes it possible to measure certain spin correlations, like $C_{xz}$.  
The signal also comes from interference between the amplitudes $B_{1/2,1/2}$ and $B_{-1/2,1/2}$.  
But the variety of spin states also dilutes quantum entanglement between the final-state baryons.  
This results in a highly mixed density matrix, which lowers the entanglement level.  
This feature has been confirmed in $\chi_{c2}$ decays into vector meson pairs \cite{Fabbrichesi:2024rec}.

Notably, both the polarization $P_y$ and the correlation tensor $C_{ij}$ depend not only on the phase $\Delta \Phi$ but also on the relative magnitude $x$ of the helicity amplitudes. 
Experimentally, the parameters $x$ and $\Delta \Phi$ can be determined through a joint angular distribution analysis of the $B\bar{B}$ decay chain. 
Theoretically, the relative magnitude $x$ can be calculated within frameworks such as the quark model or QCD-inspired approaches, which provides a first-principles prediction for the decay dynamics.
This interplay between a measurable quantum correlation pattern ($P_y$, $C_{ij}$) and the fundamental dynamical parameters ($x$, $\Delta \Phi$) makes the $\chi_{c2}$ system a unique laboratory for testing decay models. The Bell nonlocality and entanglement will be explored in Section \ref{secBell} and \ref{secCon}.

\section{Bell nonlocality in $\chi_{cJ} \to B\bar{B}$ Decays}\label{secBell}
Bell's theorem, through the derivation of the Clauser-Horne-Shimony-Holt (CHSH) inequality \cite{Clauser:1969ny}, provides a quantitative criterion to assess whether the correlations exhibited by a quantum state are compatible with local hidden variable theories. A violation of the CHSH inequality ($\mathbf{m}_{12} > 1$) signals that, under the premise of the quantum mechanics framework is accepted, the correlation predicted by the quantum state cannot be reproduced by any local hidden variable theory \cite{Horodecki:1995nsk}.

However, the test of CHSH inequality at colliders remains an open question regarding its ability to refute local realism, primarily due to fundamental differences from optical Bell test setups. Collider experiments rely on quantum state tomography to reconstruct the spin density matrix from angular distributions of decay products, rather than allowing experimenters to freely and independently choose measurement settings. This absence of tunable optical inputs introduces an element of trust in the experimental platform, meaning the results constitute a witness of Bell non-locality rather than a true, device-independent Bell test. The effectiveness and significance of this loophole are different in collider experiments from those in optical experiments, and have been revisited in Refs \cite{Fabbrichesi:2024rec, Fabbrichesi:2025aqp,Shi:2019kjf}. As reported recently in a BESIII measurement, the high significance of $\Lambda\bar\Lambda$ spin correlations has been observed, addressing the locality loophole by exploiting the weak decay distances of hyperons to achieve spacelike separation~\cite{BESIII:2025vsr}.

Nevertheless, showing that a quantum state violates a Bell inequality is still a significant physical result. It tells us that the spin correlations in a decay process cannot be explained by local realism. As a unique quantum feature, entanglement is not only widely applied in quantum science but also demonstrates novel value in high-energy particle physics: it can serve as a sensitive quantum probe in the search for evidence of CP violation \cite{Du:2024sly}. In this spirit, we do not claim a definitive loophole-free test of the CHSH inequality. Instead, we use the helicity amplitudes measured by BESIII to see if the \(\chi_{cJ} \to B\bar{B}\) system shows signs of Bell nonlocality. This gives us a concrete, data-driven foundation for planning future experiments.

To assess whether the $\chi_{cJ}\to B\bar B$ system exhibits Bell nonlocality, the experimentally measured spin correlation data are used to construct the correlation matrix $C$ in Eq.~(\ref{rbb}). The symmetric matrix \(M = CC^T\) is then formed, and its eigenvalues \(m_1 \geq m_2 \geq m_3\) are computed. According to the Horodecki condition \cite{Horodecki:1995nsk}, the Bell inequality \(I_2 \leq 2\) is violated if and only if \(m_1 + m_2 > 1\). The maximum value of the CHSH observable for a given quantum state is given by
\begin{align}
	\mathcal{B}_\text{max}[\rho] = 2 \sqrt{m_1+m_2},
\end{align}
where $m_1$ and $m_2$ are the two largest eigenvalues of the matrix $M = CC^{\mathrm{T}}$. 
A state is considered Bell nonlocal if it violates the CHSH inequality, i.e., if
\begin{align}
	\mathcal{B}[\rho] > 2.
\end{align}
The quantity $\mathbf{m}_{12} \equiv m_1 + m_2$ (with $\mathbf{m}_{12} \in [0,2]$) serves as a direct measure of the nonlocality, with the violation condition becoming $\mathbf{m}_{12} > 1$.

Using the correlation tensors $C_{ij}$ derived in Sec.~\ref{secSpinCorr}, we compute the measure $\mathbf{m}_{12}$ for each $\chi_{cJ} \to B\bar{B}$ decay.
\begin{align}\label{eq:m12}
	\mathbf{m}_{12} &= 2,~~~ \mathrm{for~ \chi_{c0}},\nonumber\\
	\mathbf{m}_{12} &= 1+\left(\frac{(2-r_1^2)\sin^2 \theta_1}{2r_1^2+(2-r_1^2)\sin^2 \theta_1}\right)^2, ~~~\mathrm{for ~\chi_{c1}},\\
	\mathbf{m}_{12} &= \text{Max}(m_1 + m_2, m_1 + m_3, m_2+ m_3),~~~ \mathrm{for ~\chi_{c2}},\nonumber
\end{align}
For $\chi_{c2}$, the eigenvalues $m_i$ (with $i=1,2,3$) of $M = C^\top C$ are functions of $\mathcal{F}_1$--$\mathcal{F}_4$ defined in Eqs.~(\ref{rbbc2}) and~(\ref{eq:F1F4}). Their explicit forms are
\begin{widetext}
	\begin{align}
	m_1 &= \frac{(\mathcal{F}_3- \mathcal{F}_4)^2}{\mathcal{F}_1^2}, \nonumber\\
	m_2 &= \frac{1}{2 \mathcal{F}_1^2}\left(\mathcal{F}_1^2+\mathcal{F}_2^2+\mathcal{F}_3^2 -4 \mathcal{F}_1 \mathcal{F}_4+2\mathcal{F}_3 \mathcal{F}_4+ 5\mathcal{F}_4^2 + \mathcal{F}_2^2 \cos 2\Delta \Phi\right. \nonumber \\
    &\quad\left.-\sqrt{(\mathcal{F}_1+\mathcal{F}_3- \mathcal{F}_4)^2 \left((-\mathcal{F}_1+ \mathcal{F}_3+ 3\mathcal{F}_4)^2+2 \mathcal{F}_2^2 \cos (2\Delta \Phi )+2 \mathcal{F}_2^2\right)}\right),\nonumber\\
	m_3 &= \frac{1}{2 \mathcal{F}_1^2}\left(\mathcal{F}_1^2+\mathcal{F}_2^2+\mathcal{F}_3^2 -4 \mathcal{F}_1 \mathcal{F}_4+2\mathcal{F}_3 \mathcal{F}_4+ 5\mathcal{F}_4^2 + \mathcal{F}_2^2 \cos 2\Delta \Phi\right. \nonumber \\
    &\quad\left.+\sqrt{(\mathcal{F}_1+\mathcal{F}_3- \mathcal{F}_4)^2 \left((-\mathcal{F}_1+ \mathcal{F}_3+ 3\mathcal{F}_4)^2+2 \mathcal{F}_2^2 \cos (2\Delta \Phi )+2 \mathcal{F}_2^2\right)}\right).
	\end{align}
\end{widetext}

The $\chi_{c0}$ decay yields maximal violation with $\mathbf{m}_{12}=2$, due to the maximum entanglement of $B\bar B$ pair, while the calculation of $\mathbf{m}_{12}$ depends on the helicity amplitude ratio in the $\chi_{c1,2}$ decays. The full numerical analysis and physical implications are presented in Sec.~\ref{NumAna}.

\section{Quantum Entanglement in $\chi_{cJ} \to B\bar{B}$ Decays}\label{secCon}
Quantum entanglement, as a form of non-classical correlation, serves as a fundamental resource for quantum information processing.
For the bipartite $B\bar{B}$ system described by the density matrix $\rho^{B\bar{B}}$, the entanglement of formation is quantified by the concurrence $\mathcal{C}[\rho]$ \cite{Wootters:1997id}.

For pure states, quantum entanglement can be quantified using the von Neumann entropy. Here, however, we uniformly adopt concurrence to measure the entanglement in the \(B\bar B\) system.  For a density matrix $\rho$, the concurrence is defined as $\mathcal{C}[\rho] = \max(0, t_1 - t_2 - t_3 - t_4)$, where $t_i$ are the square roots of the eigenvalues of the matrix $R=\rho (\sigma_y \otimes \sigma_y) \rho^* (\sigma_y \otimes \sigma_y)$ in decreasing order.
A state is separable if $\mathcal{C}[\rho]=0$ and maximally entangled if $\mathcal{C}[\rho]=1$.

Using the joint spin density matrices $\rho^{B\bar{B}}$ derived in Sec. \ref{secSpinCorr}, we calculate the concurrence for the $\chi_{cJ}$ decays.
\begin{align}\label{eq:con01}
	\mathcal{C}[\rho] & =1, ~~~\mathrm{for~ \chi_{c0}}, \nonumber \\
	\mathcal{C}[\rho] &= |\frac{(2-r_1^2)\sin^2 \theta_1}{2r_1^2+(2-r_1^2)\sin^2 \theta_1}|, ~~~\mathrm{for ~\chi_{c1}}  
\end{align}
For $\chi_{c2}$, the concurrence depends on $\theta_1$, $x$, and $\Delta\Phi$ and will be analyzed numerically in Sec.~\ref{NumAna}.

The results indicate that the \(\chi_{c0}\) state is maximally entangled, while the entanglement of \(\chi_{c1}\) is modulated, and no clear evidence of entanglement is observed in \(\chi_{c2}\) decays.

\section{Numerical Analysis and Phenomenological Discussion}\label{NumAna}
\subsection{Input parameters}
We now give a full numerical estimation, based on our earlier analysis in Secs.~\ref{secBell} and~\ref{secCon}. To move from general formulas to clear predictions, we use measured or constrained experimental values. For the $\psi(2S)\to\gamma\chi_{c1,2}$ decays,  the parameters $r_i(i=1,2,3)$ and $\Delta\Phi_j(j=2,3)$ defined in Eqs. \eqref{rc1}-\eqref{rc2} are taken as the measurements \cite{BESIII:2025gof}  $r_1=1.037 \pm 0.055,r_2=1.48 \pm0.148, r_3=1.85\pm0.180 $, and $\Delta\Phi_1=0.042\pm0.112, \Delta\Phi_2=0.37\pm0.21,\Delta\Phi_3=0.13 \pm 0.13$ in unit of rad.  For $\chi_{c2}\to B\bar B$ decays, the amplitude ratio $x$ is extracted using the baryon angular distribution parameters according to the Eqs. \eqref{eq:angdis}.  The $x$ values are given in Table~\eqref{tab:AngularPars} .

\begin{table}[htbp!]
\centering
\caption{Summary of angular distribution parameters in $\chi_{c2} \to B\bar{B}$ Decays}
\label{tab:AngularPars}
\resizebox{0.48\textwidth}{!}{
\begin{tabular}{ccccc}
\hline
Decays & $\alpha_{J}$ & $x$ & $\Delta \Phi$ & Reference \\ \hline
$\chi_{c2}\to p \bar{p}$ 
  & $-0.26 \pm 0.17$ 
  & $x=1.31\pm1.15$ 
  & $\pi/2$ (fixed) 
  & \cite{BESIII:2013nam} \\ \hline
$\chi_{c2}\to \Lambda \bar{\Lambda}$  
  & $-0.211 \pm 0.112$ 
  & $x=1.74 \pm 0.15$ 
  & $-0.37 \pm 0.16$ 
  & \cite{BESIII:2025gof} \\ \hline
$\chi_{c2} \to \Xi^{0} \bar{\Xi}^{0}$ 
  & $-0.65 \pm 0.38$ 
  &  $x\in (0, 1.25)$ 
  & $\pi/2$ (fixed) 
  & \multirow{2}{*}{\cite{BESIII:2022mfx}} \\ \cline{1-4}
$\chi_{c2} \to \Xi^{-} \bar{\Xi}^{+}$ 
  & $-0.34 \pm 0.34$ 
  & $x=0.88^{+1.51}_{-0.88}$
  & $\pi/2$ (fixed) 
  & \\ \hline
\end{tabular}}
\end{table}

\subsection{Profile of Bell nonlocality}\label{BellNon}
We employ the Horodecki condition \(\mathbf{m}_{12}\) to quantify Bell nonlocality in the decays of \(\chi_{cJ}\) states. The \(\chi_{c0}\) state yields a constant value of \(\mathbf{m}_{12}=2\), corresponding to a maximal and uniform violation of the Bell inequality across all angles.

\begin{figure}[htbp]  
    \centering  \includegraphics[width=0.4\textwidth]{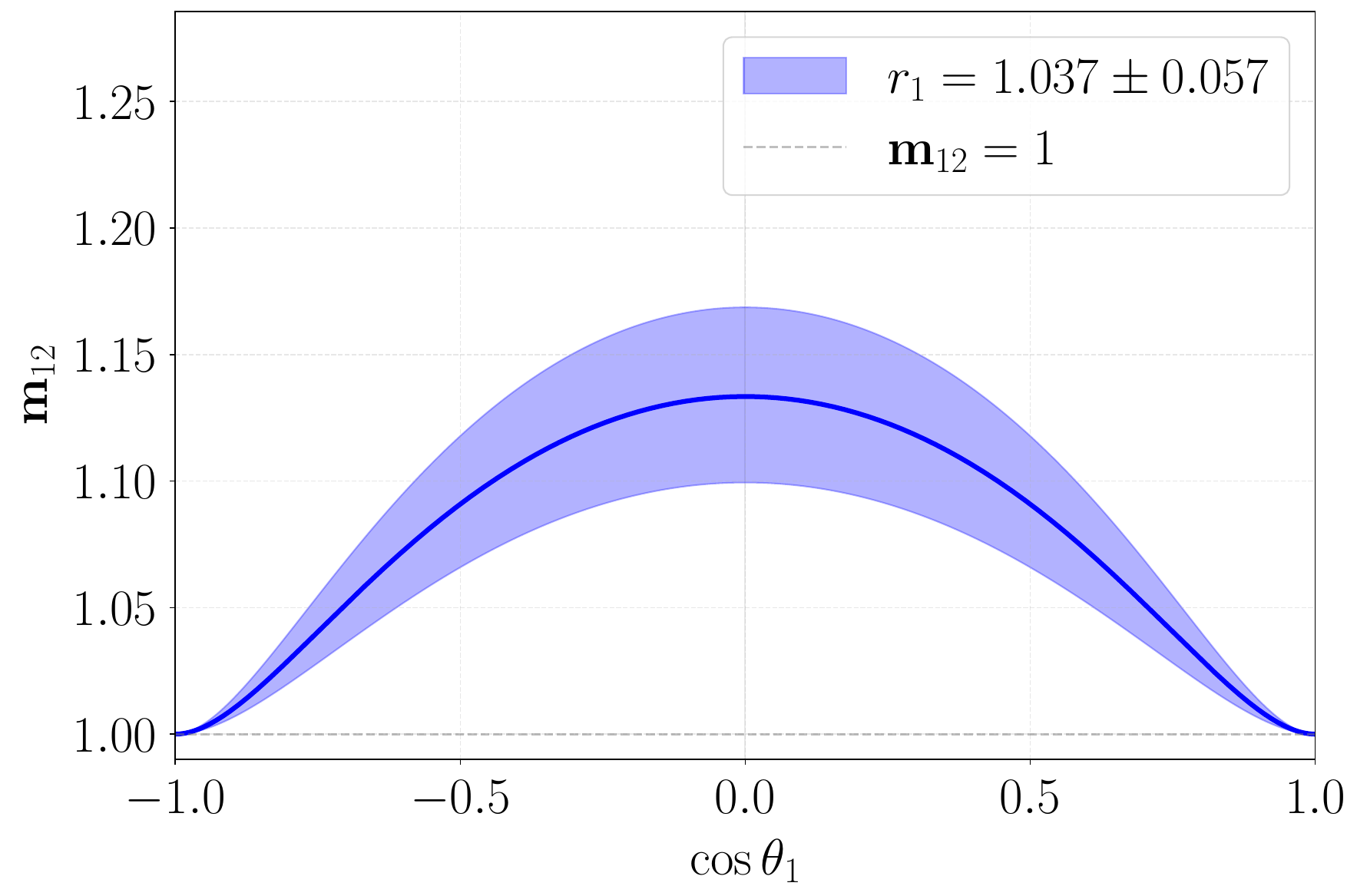}
    \caption{The Horodecki condition $\mathbf{m}_{12}$ as functions of $\cos \theta_{1}$ of the baryon in $e^{+}e^{-} \to \psi(2S) \to \gamma \chi_{c1}, \chi_{c1} \to B\bar{B}$ decays. The parameter $r_1$ is fixed at $1.307\pm 0.057$}  
    \label{fig:Bellc01}  
\end{figure}

Figure \ref{fig:Bellc01} shows the distribution of \(\mathbf{m}_{12}\) as a function of the baryon helicity angle \(\cos\theta_1\). The distribution peaks at \(\cos\theta_1=0\), with the shaded band representing the uncertainty from the parameter \(r_1\). A \(\chi^2\) estimation yields a significance of \(2.7\sigma\) for the hypothesis \(\mathbf{m}_{12}>1\) at $\theta=\pi/2$, indicating that the observed Bell inequality violation is not statistically significant. From Eqs. (\ref{eq:m12}) and (\ref{eq:con01}), one has a relation between the $\mathbf{m}_{12}$ and quantum concurrence $\mathcal{C}$ for $B\bar B$  system in $\chi_{c0,1}$ decays:
    \begin{equation}
        \mathbf{m}_{12} = 1+ \mathcal{C}^2.
    \end{equation}

\begin{figure}
    \centering
    \includegraphics[width=0.9\linewidth]{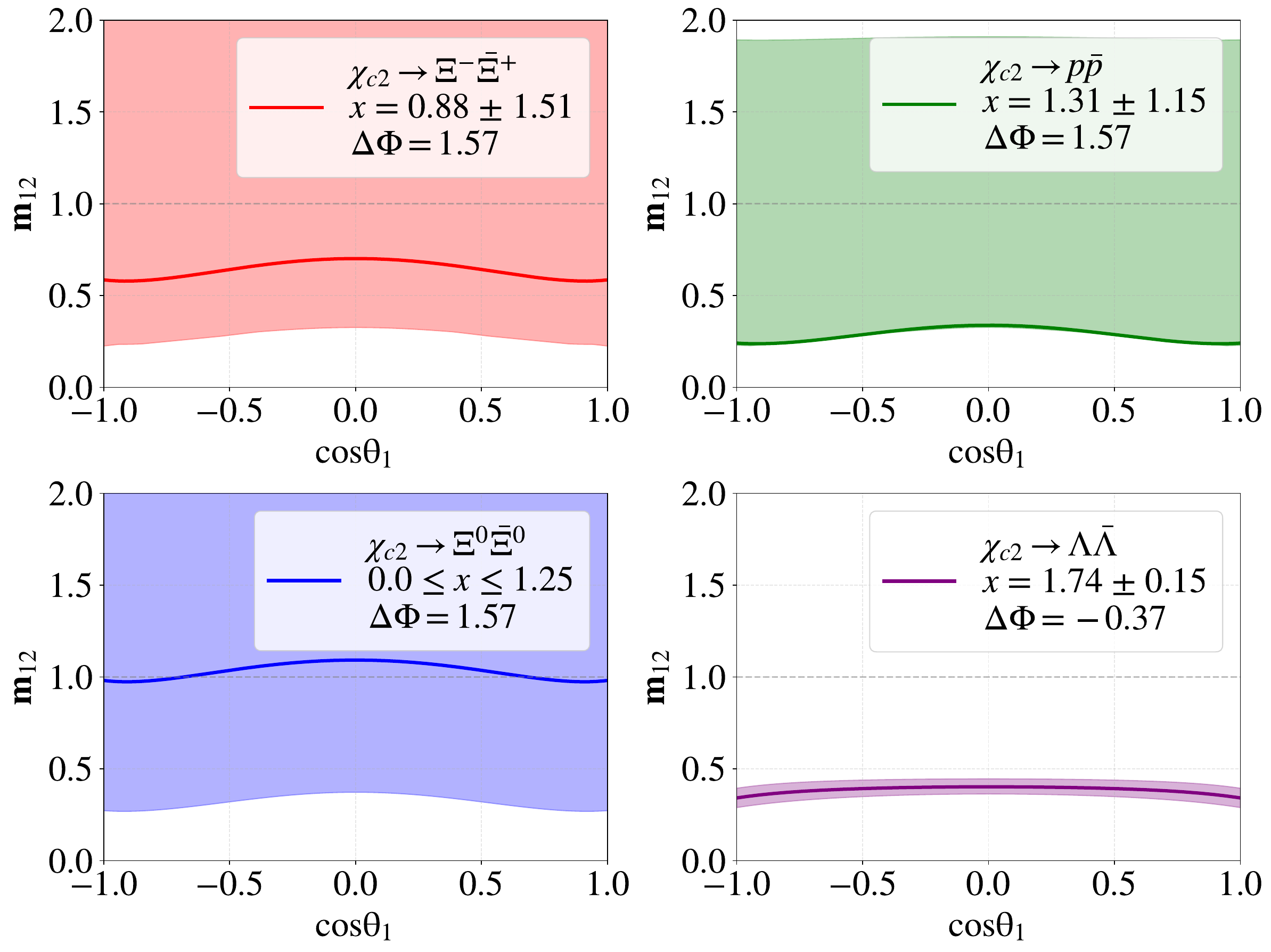}
    \caption{The Horodecki condition $\mathbf{m}_{12}$ as functions of $\cos \theta_1$ in $\chi_{c2} \to B\bar{B}$ decays with $r2, r3$ fixed to the measurements \cite{BESIII:2025gof}. The line is calculated with the center value of $x$.}
    \label{fig:m12chic2BB_theta_1}
\end{figure}

Figure \ref{fig:m12chic2BB_theta_1} presents the Horodecki condition $\mathbf{m}_{12}$ distribution as a function of $\cos\theta_1$ for $\chi_{c2}$ decays into different baryon-antibaryon final states. The solid line represents the $\mathbf{m}_{12}$ values calculated using the extracted parameters $x$, while the shaded band indicates the uncertainty originating from the errors on $x$. Apart from the $\Delta\Phi$ accessible in $\chi_{c2}\to \Lambda\bar\Lambda$ decays, the relative phase has not been measured in other modes. Therefore, for the $\Xi^{0}\bar\Xi^{0}$, $\Xi^{-}\bar\Xi^{+}$, and $p\bar p$ channels, $\Delta\Phi$ is set to $\pi/2$.   For the $\chi_{c2}\to\Lambda\bar\Lambda$ decay, the condition $\mathbf{m}_{12}<1$ is significantly observed, indicating no violation of the Bell inequality. For other decay channels, we also find no evidence of Bell inequality violation, due to the large uncertainties in the measured $x$ parameters.

\subsection{Profile of quantum entanglement}

\begin{figure}[htbp]  
    \centering  
    \includegraphics[width=0.4\textwidth]{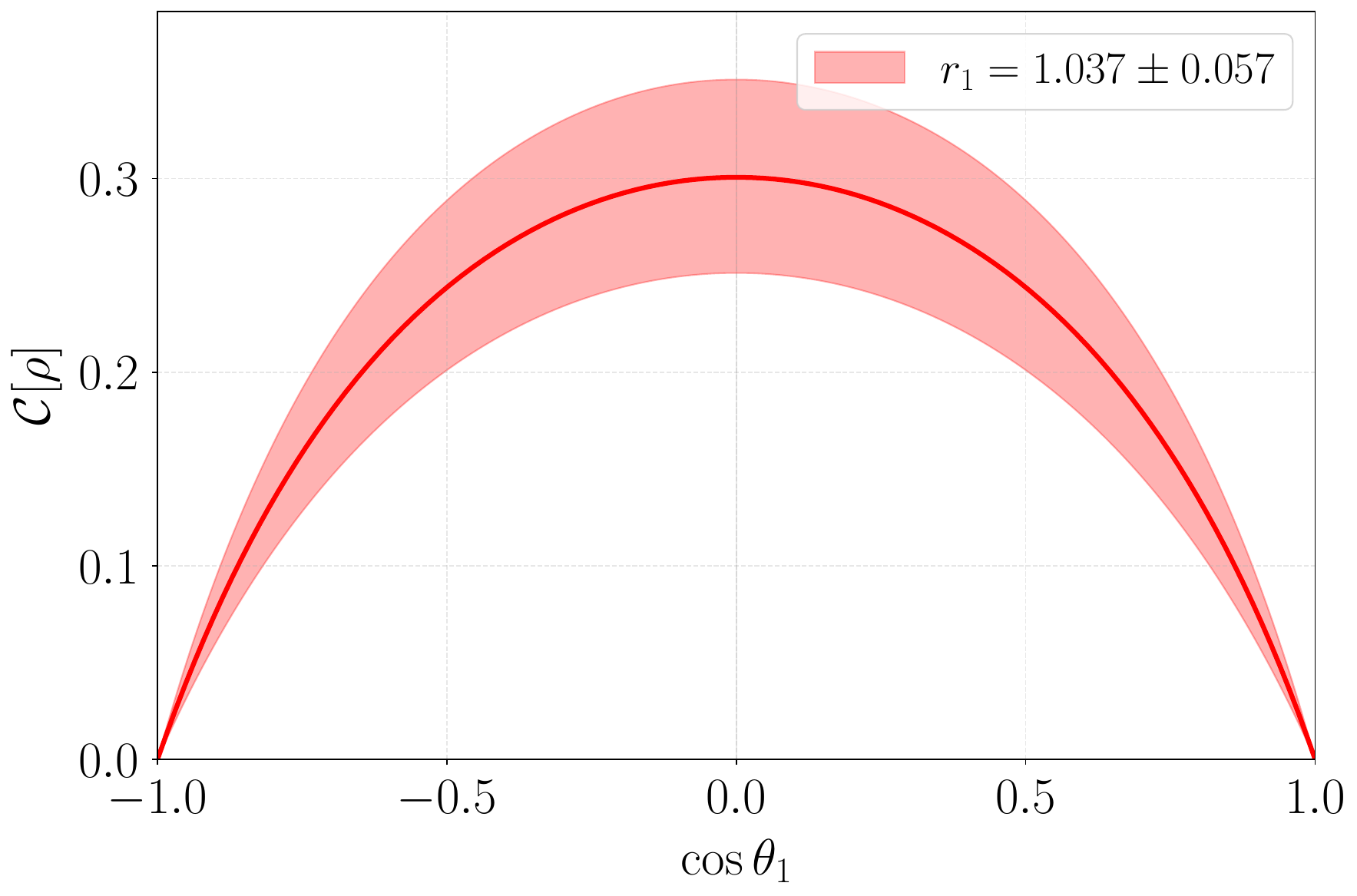}
    \caption{The concurrence $\mathcal{C}[\rho]$ as functions of $\cos \theta_{1}$ in $e^{+}e^{-} \to \psi(2S) \to \gamma \chi_{c1}, \chi_{c1} \to B\bar{B}$ decays.}  
    \label{fig:conc1_noE1}  
\end{figure}

We employ concurrence, \(\mathcal{C}[\rho]\), to quantify the entanglement of the \(B\bar{B}\) system in \(\chi_{cJ}\) decays. For the pure \(\chi_{c0}\) state, the concurrence reaches its maximum value of \(\mathcal{C}[\rho] = 1\), which equals the von Neumann entropy after tracing out one subsystem from the density matrix \(\rho^{B\bar{B}}\).  Figure \ref{fig:conc1_noE1} shows the angular dependence of  concurrence of the $B\bar B$ states produced from $\chi_{c1}$ decays.   It has a bell-like shape, with maximum $\sim0.3 $ at $\theta=\pi/2$, 

\begin{figure}[htbp]  
    \centering  
    \includegraphics[width=0.4\textwidth]{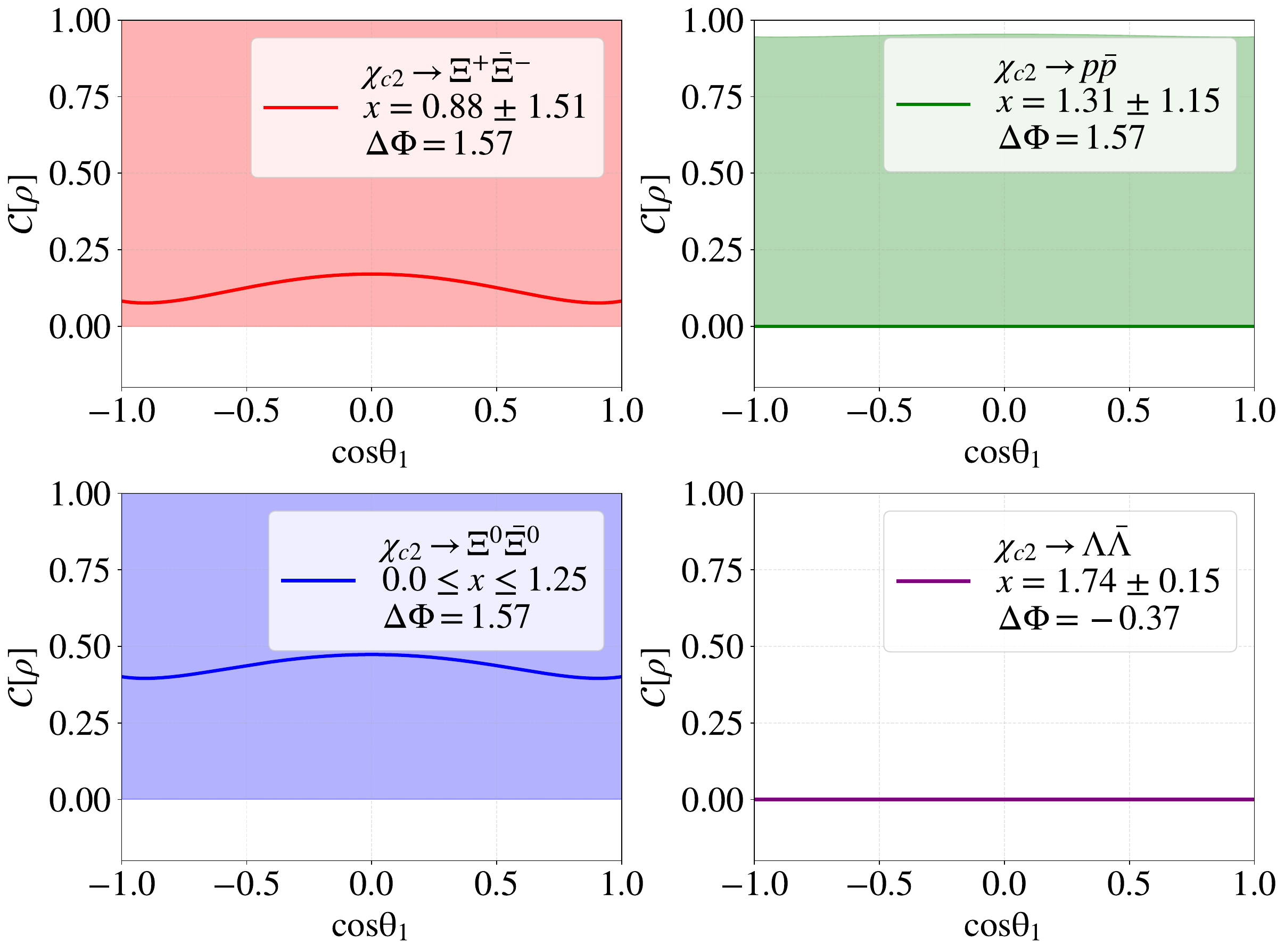}
    \caption{The concurrence $\mathcal{C}[\rho]$ as functions of $\cos \theta_{1}$ in $e^{+}e^{-} \to \psi(2S) \to \gamma \chi_{c2}, \chi_{c2} \to B\bar{B}$ decays. The line is calculated with the center values of $x$.}  
    \label{fig:conc2}  
\end{figure}

Figure~\ref{fig:conc2} shows the concurrence for the $B\bar B$ states from $\chi_{c2}$ decays. With the extracted $x$ values, we find no indication of concurrence for the $p\bar p$, $\Xi^{0}\bar \Xi^{0},\Xi^{-}\bar \Xi^{+}$ and $\Lambda\bar \Lambda$ states. 

For the $\chi_{c2}$ decays, the Horodecki parameter satisfies $\mathbf{m}_{12} < 1$. Therefore, no violation of the Bell inequality is observed in these channels. Instead, one may examine the entanglement content of the $B\bar{B}$ pair via negativity, and it can be shown that the pair is separable. The partial transposition of $\rho^{\bar{B}}$ can be obtained by transposing the $2\times2$ sub‑matrix of $\rho^{B\bar{B}}$; if all eigenvalues of $\rho^{\bar{B}}$ are non‑negative, the $B\bar{B}$ pair is in a separable state \cite{vidal}. Using the measured parameters for $\chi_{c2}$ production, the separable condition is found to be \(1.045 \le x \le 2.231\). Within the 1$\sigma$ range of the estimated parameter \(x\), all values satisfy this separable condition.

\subsection{Feasibility and Prospects}\label{ExpFeas}
Based on the numerical results obtained from current experimental inputs, we now assess the feasibility of observing the predicted quantum correlations. This assessment takes into account the existing BESIII data sets and other facilities.

\subsubsection{Current Data and Future Precision}

BESIII has accumulated the world's largest sample of $\psi(2S)$ events. For $\psi(2S) \to \gamma \chi_{cJ},~\chi_{cJ} \to \Lambda\bar{\Lambda}$ decay, BESIII has recently published a high-statistics partial wave analysis using the full $\psi(2S)$ data sample of $2.712\times10^9$ events \cite{BESIII:2025gof}. Both the helicity amplitude ratios and the relative phase angles are measured with high precision, and systematic uncertainties dominate the error budget (e.g., $R_{\chi_{c2}}=0.575\pm0.048\pm0.018$, $\Delta\Phi_{\chi_{c2}}=-0.37\pm0.15\pm0.05$~rad). Consequently, the extracted $\mathbf{m}_{12}$ and concurrence for this channel are sufficiently precise to conclusively verify that the Horodecki condition is satisfied.

For $\chi_{cJ} \to p\bar{p}$ and $\chi_{cJ} \to \Xi\bar{\Xi}$ decays, the BESIII measurements are only based on partial data sets ($p\bar{p}$: $1.06\times10^8$ $\psi(2S)$ events \cite{BESIII:2013nam}, $\Xi^{-}\bar{\Xi}^{+}$, $\Xi^{0}\bar{\Xi}^{0}$: $4.481\times10^8$ $\psi(2S)$ events \cite{BESIII:2022mfx}). The branching fractions of these decay modes are of the same order as that of $\chi_{cJ} \to \Lambda\bar{\Lambda}$ (typically $\mathcal{O}(10^{-4} \sim 10^{-5})$). Therefore, with the full $2.712\times10^9$ $\psi(2S)$ data sample available at BESIII, the signal yields for $p\bar{p}$ and $\Xi\bar{\Xi}$ as well as $\Sigma \bar{\Sigma}$ final states are expected to be comparable to those of the $\Lambda\bar{\Lambda}$ channel. By scaling the systematic uncertainties from the $\Lambda\bar{\Lambda}$ measurement, we estimate that the precision of helicity parameters for these channels could be improved by a factor of $2$--$3$ once the full data set is analyzed. This would significantly reduce the error bands as shown in the figures of these decay modes, and enable to draw definitive conclusions on their Bell nonlocality and entanglement properties.

\subsubsection{Prospects at other facilities}

Building on the above analysis, we next examine the feasibility of investigating $\chi_{cJ} \to B\bar{B}$ decays at other high-energy collider experiments. 

LHCb and the Belle experiments offer valuable opportunities for heavy-flavor physics, but two challenges are confronted to probe $\chi_{cJ} \to B\bar{B}$ decays: low signal yield and the complexity of reconstructing the spin density matrix.

At LHCb, $\chi_{cJ}$ states are predominantly produced via $b$-hadron decays ($B \to \chi_{cJ} K$) \cite{LHCb:2013mbz} or prompt production in $pp$ collisions \cite{LHCb:2013ofo}, but the yield is less than a tenth of that at BESIII. At the Belle experiments, which operates at the $\Upsilon(4S)$ resonance \cite{Belle-II:2018jsg, Belle-II:2010dht}, the production rate via an initial state radiation (ISR) process, $e^+e^- \to \gamma_{ISR}\psi(2S)$, and $\psi(2S) \to \gamma \chi_{cJ}$ is also severely suppressed, with yields orders of magnitude below BESIII's $2.7\times10^9$ sample \cite{Belle:2015hcs}. The both experiments face a fundamental challenge to reconstruct the  $\chi_{cJ}$ spin density matrix. Compared to $e^+e^-$ annihilation at the BESIII, the $\chi_{cJ}$ mesons may be abundantly produced via other mechanisms at both facilities, but their spin density matrices are not directly accessible.  One has to do a partial wave analysis to extract the helicity amplitudes of $\chi_{cJ}$ states by considering different production mechanisms. Nevertheless, if these matrices can be experimentally reconstructed, the same quantum tomography framework remains applicable.

The proposed Super $\tau$-Charm Factory (STCF) in China is designed to operate in the $\tau$-charm energy region with a peak luminosity of $0.5\times10^{35}~\text{cm}^{-2}\text{s}^{-1}$ --- two orders of magnitude higher than BEPCII \cite{Achasov:2023gey}. With such a facility, the production rate of $\psi(2S) \to \gamma \chi_{cJ}$ events would be increased by a factor of $\sim100$. With sufficiently high statistics, it would reduce statistical uncertainties to the sub-percent level, and the measurement precision would be dominated by systematic uncertainties, enabling percent-level tests of quantum correlation and measurements of helicity parameters for all ground baryonic pairs. If a polarized electron beam is used at the STCF in the future, the hyperons will acquire a longitudinal polarization. This will significantly enhance the spin analysis power for hyperons \cite{Salone,Zeng}, thereby providing more favorable conditions for studying the polarization correlation of baryon pairs and for testing Bell's inequality with polarized beams \cite{Zhang:2026nwm}. This would address the fundamental limitations discussed in Sec.~\ref{secBell}. Therefore, the STCF represents the definitive future facility for the kind of quantum foundation studies advocated in this work.

\section{Summary and Outlook}
In this work, we have presented a comprehensive theoretical analysis of Bell nonlocality and quantum entanglement in the decays $\chi_{cJ} \to B\bar{B}$ ($J=0,1,2$), specifically for $\chi_{cJ}$ mesons produced via the radiative transition $\psi(2S) \to \gamma \chi_{cJ}$ in $e^+e^-$ annihilation. By deriving the joint spin density matrix for the baryon-antibaryon system within the helicity formalism, we computed the complete set of polarization and spin correlation parameters. From these, we obtained analytical formulas for the Bell measure $\mathbf{m}_{12}$ and the concurrence $\mathcal{C}[\rho]$, revealing a striking and hierarchical picture of quantum correlations across the $\chi_{cJ}$ family under this well-defined production condition.

The $\chi_{c0}$ decay yields a $B\bar{B}$ pair in a pure, maximally entangled spin-triplet state ($\mathcal{C}[\rho]=1$), which guarantees a maximal violation of the Bell inequality ($\mathbf{m}_{12}=2$). In contrast, the $\chi_{c1}$ decay produces a state with persistent, yet non-maximal, entanglement and Bell nonlocality that depends on the angular distribution; violation is observed over a wide angular range and disappears only exactly in the forward and backward directions, where transverse spin correlations are quenched. Within the uncertainty of the estimated helicity parameters for $\chi_{c2} \to B\bar{B}$, we find that the $B\bar{B}$ pair is in a separable state, and no indication of Bell inequality violation is observed. Using the Horodecki criterion and the quantum concurrence, we show their dependence on the helicity ratio $x$ as presented in Table~\ref{tab:AngularPars}.

\begin{table}[htbp!]
\centering
\caption{Entanglement and Bell nonlocality in $\chi_{c2} \to B\bar{B}$ decays as functions of $x$}
\label{tab:AngularPars}
\resizebox{0.48\textwidth}{!}{
\begin{tabular}{ccc}
\hline
$x$ range & Concurrence $\mathcal{C}$ & CHSH violation \\ 
\hline
$x = 0$ & $\mathcal{C} = 1$ & Violated ($\mathcal{B}=2\sqrt{2}$)  \\ 
$0 < x < 0.678$ & $0 < \mathcal{C} < 1$ & Violated ($\mathcal{B}>2$)  \\
$0.678 \leq x < 1.045$ & $0 < \mathcal{C} < 1$ & Not violated ($\mathcal{B} \leq 2$)  \\
$1.045 \leq x \leq 2.231$ & $\mathcal{C} = 0$ & Not violated  \\
$2.231 < x \leq 8.969$ & $0 < \mathcal{C} < 1$ & Not violated  \\
$x > 8.969$ & $0 < \mathcal{C} < 1$ & Violated ($\mathcal{B} > 2$)  \\
\hline
\end{tabular}}
\end{table}

The predictions in this work are specific to the production chain $e^+e^- \to \psi(2S) \to \gamma \chi_{cJ}$. They can be tested directly at BESIII using its full $\psi(2S)$ data sample.
Looking forward, this analysis provides clear, measurable benchmarks for experiments at BESIII and future facilities such as the Super Tau-Charm Factory. In particular, the predicted angular distributions for $\chi_{c2}$ decays offer a direct means to extract the parameters $x$ and $\Delta\Phi$ more accurately. Precise measurement of these parameters will help infer quantum correlations via the mapping established in this work. A detailed assessment of the experimental feasibility, including the required event yields and the prospects at BESIII, LHCb, the Belle experiments, and future STCF, is also presented (see~\ref{ExpFeas}). These indicate that the $\chi_{cJ}$ system produced via this radiative transition serves as a calibrated laboratory at the intersection of quantum information science and high-energy physics. Furthermore, studying quantum correlations in charmonium decays opens a new window into foundational quantum mechanics, enabling tests at energy scales and in regimes that are far removed from traditional optical experiments.

\section{acknowledgements}
The work is partly supported by the National Natural Science Foundation of China(NSFC) under Grants No. 12575112 and Program of Science and Technology Development Plan of Jilin Province of China under Contract No. 20230101021JC.



\begin{thebibliography}{**}






	\bibitem{Horodecki:2009zz}
	R.~Horodecki, P.~Horodecki, M.~Horodecki and K.~Horodecki,
	\href{doi:10.1103/RevModPhys.81.865}{Rev. Mod. Phys. \textbf{81}, 865-942 (2009)}
	[arXiv:quant-ph/0702225 [quant-ph]].

	\bibitem{Bell:1964kc}
	J.~S.~Bell,
	\href{doi:10.1103/PhysicsPhysiqueFizika.1.195}{Physics Physique Fizika \textbf{1}, 195-200 (1964)}

\bibitem{Clauser:1969ny}
J.~F.~Clauser, M.~A.~Horne, A.~Shimony and R.~A.~Holt,
\href{doi:10.1103/PhysRevLett.23.880}{Phys. Rev. Lett. \textbf{23}, 880-884 (1969)}

\bibitem{Brunner:2013est}
N.~Brunner, D.~Cavalcanti, S.~Pironio, V.~Scarani and S.~Wehner,
\href{doi:10.1103/RevModPhys.86.419}{Rev. Mod. Phys. \textbf{86}, 419 (2014)}
[arXiv:1303.2849 [quant-ph]].

\bibitem{Freedman:1972zza}
S.~J.~Freedman and J.~F.~Clauser,
\href{doi:10.1103/PhysRevLett.28.938}{Phys. Rev. Lett. \textbf{28}, 938-941 (1972)}

\bibitem{Aspect:1981nv}
A.~Aspect, P.~Grangier and G.~Roger,
\href{doi:10.1103/PhysRevLett.49.91}{Phys. Rev. Lett. \textbf{49}, 91-97 (1982)}

\bibitem{Aspect:1982fx}
A.~Aspect, J.~Dalibard and G.~Roger,
\href{doi:10.1103/PhysRevLett.49.1804}{Phys. Rev. Lett. \textbf{49}, 1804-1807 (1982)}

\bibitem{Hagley:1997bob}
E.~Hagley, X.~Ma{\^\i}tre, G.~Nogues, C.~Wunderlich, M.~Brune, J.~M.~Raimond and S.~Haroche,
\href{doi:10.1103/PhysRevLett.79.1}{Phys. Rev. Lett. \textbf{79}, no.1, 1-5 (1997)}

\bibitem{Fabbrichesi:2021npl}
M.~Fabbrichesi, R.~Floreanini and G.~Panizzo,
\href{doi:10.1103/PhysRevLett.127.161801}{Phys. Rev. Lett. \textbf{127}, no.16, 16 (2021)}
[arXiv:2102.11883 [hep-ph]].

\bibitem{Bernal:2024xhm}
A.~Bernal, P.~Caban and J.~Rembieli{\'n}ski,
\href{doi:10.1038/s41598-025-07747-3}{Sci. Rep. \textbf{15}, no.1, 23410 (2025)}
[arXiv:2405.16525 [hep-ph]].


\bibitem{ATLAS:2023fsd}
G.~Aad \textit{et al.} [ATLAS],
\href{doi:10.1038/s41586-024-07824-z}{Nature \textbf{633}, no.8030, 542-547 (2024)}
[arXiv:2311.07288 [hep-ex]].

\bibitem{Ehataht:2023zzt}
K.~Ehat{\"a}ht, M.~Fabbrichesi, L.~Marzola and C.~Veelken,
\href{doi:10.1103/PhysRevD.109.032005}{Phys. Rev. D \textbf{109}, no.3, 032005 (2024)}
[arXiv:2311.17555 [hep-ph]].

\bibitem{BESIII:2018cnd}
M.~Ablikim \textit{et al.} [BESIII],
\href{doi:10.1038/s41567-019-0494-8}{Nature Phys. \textbf{15}, 631-634 (2019)}
[arXiv:1808.08917 [hep-ex]].

\bibitem{BESIII:2025vsr}
M.~Ablikim \textit{et al.} [BESIII],
\href{doi:10.1038/s41467-025-59498-4}{Nature Commun. \textbf{16}, 4948 (2025)}
[arXiv:2505.14988 [hep-ex]].

\bibitem{Chung:1971ri}
S.~U.~Chung,
\href{doi:10.5170/CERN-1971-008}{CERN Yellow Reports: Monographs}

\bibitem{Wu:2024asu}
S.~Wu, C.~Qian, Q.~Wang and X.~R.~Zhou,
\href{doi:10.1103/PhysRevD.110.054012}{Phys. Rev. D \textbf{110}, no.5, 054012 (2024)}
[arXiv:2406.16298 [hep-ph]].

\bibitem{Fabbrichesi:2024rec}
M.~Fabbrichesi, R.~Floreanini, E.~Gabrielli and L.~Marzola,
\href{doi:10.1103/PhysRevD.110.053008}{Phys. Rev. D \textbf{110}, no.5, 053008 (2024)}
[arXiv:2406.17772 [hep-ph]].

\bibitem{Karl:1975jp}
G.~Karl, S.~Meshkov and J.~L.~Rosner,
\href{doi:10.1103/PhysRevD.13.1203}{Phys. Rev. D \textbf{13}, 1203 (1976)}

\bibitem{BESIII:2025gof}
M.~Ablikim \textit{et al.} [BESIII],
\href{https://arxiv.org/pdf/2509.00289}{[arXiv:2509.00289 [hep-ex]].}

\bibitem{Tabakin:1985yv} 
Phys. Rev. C \textbf{31} , 1857 (1985) 

\bibitem{Horodecki:1995nsk}
R.~Horodecki, P.~Horodecki and M.~Horodecki,
\href{doi:10.1016/0375-9601(95)00214-N}{Phys. Lett. A \textbf{200}, no.5, 340-344 (1995)}


\bibitem{Fabbrichesi:2025aqp}
M.~Fabbrichesi, R.~Floreanini and L.~Marzola,
Found. Phys. \textbf{55}, no.6, 83 (2025)
doi:10.1007/s10701-025-00894-7
[arXiv:2503.18535 [quant-ph]].

\bibitem{Shi:2019kjf}
Y.~Shi and J.~C.~Yang,
\href{doi:10.1140/epjc/s10052-020-7684-5}{Eur. Phys. J. C \textbf{80}, no.2, 116 (2020)}
[arXiv:1912.04111 [hep-ph]].

\bibitem{Du:2024sly}
Y.~Du, X.~G.~He, C.~W.~Liu and J.~P.~Ma,
Eur. Phys. J. C \textbf{85}, no.11, 1255 (2025)
doi:10.1140/epjc/s10052-025-15003-1
[arXiv:2409.15418 [hep-ph]].

\bibitem{Wootters:1997id}
W.~K.~Wootters,
\href{doi:10.1103/PhysRevLett.80.2245}{Phys. Rev. Lett. \textbf{80}, 2245-2248 (1998)}
[arXiv:quant-ph/9709029 [quant-ph]].

\bibitem{BESIII:2013nam}
M.~Ablikim \textit{et al.} [BESIII],
\href{doi:10.1103/PhysRevD.88.112001}{Phys. Rev. D \textbf{88}, no.11, 112001 (2013)}
[arXiv:1310.6099 [hep-ex]].



\bibitem{BESIII:2022mfx}
M.~Ablikim \textit{et al.} [BESIII],
\href{doi:10.1007/JHEP06(2022)074}{JHEP \textbf{06}, 74 (2022)}
[arXiv:2202.08058 [hep-ex]].

\bibitem{vidal}
G. Vidal and R. F. Werner, Computable measure of entanglement, Phys. Rev. A 65, 032314 (2002).

\bibitem{LHCb:2013mbz}
R.~Aaij \textit{et al.} [LHCb],
\href{doi:10.1016/j.nuclphysb.2013.06.005}{Nucl. Phys. B \textbf{874}, 663-678 (2013)}

[arXiv:1305.6511 [hep-ex]].

\bibitem{LHCb:2013ofo}
R.~Aaij \textit{et al.} [LHCb],
\href{doi:10.1007/JHEP10(2013)115}{JHEP \textbf{10}, 115 (2013)}
[arXiv:1307.4285 [hep-ex]].


\bibitem{LHCb:2008vvz}
A.~A.~Alves, Jr. \textit{et al.} [LHCb],
\href{doi:10.1088/1748-0221/3/08/S08005}{JINST \textbf{3}, S08005 (2008)}

\bibitem{Belle-II:2018jsg}
E.~Kou \textit{et al.} [Belle-II],

\href{doi:10.1093/ptep/ptz106}{[erratum: PTEP \textbf{2020}, no.2, 029201 (2020)]}
[arXiv:1808.10567 [hep-ex]].

\bibitem{Belle-II:2010dht}
T.~Abe \textit{et al.} [Belle-II],
\href{https://arxiv.org/pdf/1011.0352}{[arXiv:1011.0352 [physics.ins-det]].}


\bibitem{Belle:2015hcs}
Y.~L.~Han \textit{et al.} [Belle],
\href{doi:10.1103/PhysRevD.92.012011}{Phys. Rev. D \textbf{92}, no.1, 012011 (2015)}
[arXiv:1506.05229 [hep-ex]].

\bibitem{Achasov:2023gey}
M.~Achasov, X.~C.~Ai, R.~Aliberti, L.~P.~An, Q.~An, X.~Z.~Bai, Y.~Bai, O.~Bakina, A.~Barnyakov and V.~Blinov, \textit{et al.}
\href{doi:10.1007/s11467-023-1333-z}{Front. Phys. (Beijing) \textbf{19}, no.1, 14701 (2024)}

[arXiv:2303.15790 [hep-ex]].
\bibitem{Salone}N. Salone, P. Adlarson, V. Batozskaya, A. Kupsc, S.Leupold, and J. Tandean, Phys. Rev. D \textbf{105}, 116022 (2022).
\bibitem{Zeng} S. Zeng, Y. Xu, X. R. Zhou, J. J. Qin, and B. Zheng, Chin.Phys. C \textbf{47}, 113001 (2023).

\bibitem{Zhang:2026nwm}
H.~W.~Zhang, X.~Cao and T.~F.~Feng,
[arXiv:2602.10389 [hep-ph]].
\end{thebibliography}
\end{document}